\documentclass[conference]{IEEEtran}
\IEEEoverridecommandlockouts

\usepackage{cite}
\usepackage{amsmath,amssymb,amsfonts}
\usepackage{algorithmic}
\usepackage{graphicx}
\usepackage{textcomp}
\usepackage{xcolor}
\usepackage{float}
\def\BibTeX{{\rm B\kern-.05em{\sc i\kern-.025em b}\kern-.08em
    T\kern-.1667em\lower.7ex\hbox{E}\kern-.125emX}}
\begin{document}

\title{Adversarial-Resilient RF Fingerprinting: A CNN-GAN Framework for Rogue Transmitter Detection
}

\author{\IEEEauthorblockN{Raju Dhakal}
\IEEEauthorblockA{
\textit{Embry-Riddle Aeronautical University}\\
Daytona Beach, FL, USA \\
dhakalr@my.erau.edu}

\and
\IEEEauthorblockN{Prashant Shekhar}
\IEEEauthorblockA{
\textit{Embry-Riddle Aeronautical University}\\
Daytona Beach, FL, USA \\
shekharp@erau.edu}

\and
\IEEEauthorblockN{Laxima Niure Kandel}
\IEEEauthorblockA{
\textit{Embry-Riddle Aeronautical University}\\
Daytona Beach, FL, USA \\
niurekal@erau.edu}}
\maketitle
\begin{abstract}
Radio Frequency Fingerprinting (RFF) has evolved as an effective solution for authenticating devices by leveraging the unique imperfections in hardware components involved in the signal generation process. In this work, we propose a Convolutional Neural Network (CNN)-based framework for detecting rogue devices and identifying genuine ones using softmax probability thresholding. We emulate an attack scenario in which adversaries attempt to mimic the RF characteristics of genuine devices by training a Generative Adversarial Network (GAN) using In-phase and Quadrature (I/Q) samples from genuine devices. The proposed approach is verified using I/Q samples collected from ten different ADALM-PLUTO Software Defined Radios (SDRs), with seven devices considered genuine, two as rogue, and one used for validation to determine the threshold. The GAN’s ability to generate realistic synthetic samples is validated by plotting the I/Q constellations of both real and synthetic signals.

\end{abstract}

\section{Introduction}

The wireless communication technologies have been growing rapidly. This growth has brought several benefits, including enhanced connectivity, increased data transmission capabilities, and the use of wireless devices in a broader range of areas, such as the Internet of Things (IoT), sensor networks, Unmanned Aerial Vehicles (UAVs), and many more \cite{en16031349}. Although they are widely used in diverse application areas, they also pose significant security threats, particularly in authenticating and managing access to the network due to the growing number of heterogeneous and potentially untrusted devices \cite{9946859}. Traditional cryptographic techniques often impose computational overhead in resource-constrained environments, such as IoT, sensor networks, and UAV networks. Recent studies have explored lightweight authentication approaches suitable for resource-constrained environments \cite{electronics11182921}. Among them, Radio Frequency Fingerprinting (RFF) has become a promising solution for authenticating wireless devices. The RFF-based approach leverages the fact that every wireless device has a unique identifier in its transmitted signal due to imperfections in the hardware components involved in signal generation \cite{dhakal2025physical}. A general RFF technique uses machine learning–based approaches to process the I/Q data from wireless devices, extract the minute variations inherent in the I/Q samples generated by different devices, and classify them accordingly. For example, Huang et al. in \cite{10433427} utilize machine learning models, such as CNN, to classify devices using RFF. Here, the authentication challenge is the presence of rogue devices that try to access the wireless network (which can be an IoT network, sensor network, or UAV network). Hence, the RFF approach must first differentiate between genuine and rogue devices and then classify the genuine devices. In this regard, various studies have been conducted to detect rogue devices and identify genuine ones \cite{s20041213, 9277909}. In an attack scenario, adversaries may not always use rogue devices directly. Sometimes, they may try to mimic the characteristics of genuine devices to gain access to the network. To address these authentication challenges, we emulate this type of scenario by generating synthetic I/Q samples that mimic the characteristics of genuine devices using Generative Adversarial Networks (GANs). We train the GAN using RF samples from genuine devices, where the generator learns to produce I/Q samples similar to those of a genuine device.
\par In this article, we propose utilizing a Convolutional Neural Network (CNN), which can distinguish between genuine and rogue devices and identify genuine devices based on the softmax probability vector generated for each sample. We evaluate our proposed approach using a test set that includes real genuine devices, real rogue devices, and synthetic rogue samples generated by the trained GAN. By doing so, we address the problem of identifying adversaries that attempt to mimic the RF characteristics of genuine devices. To the best of our knowledge, this is the first work to verify the performance of the designed approach for detecting out-of-distribution devices using I/Q samples from real devices, as well as synthetically generated samples from a GAN. 

\section{Related works}\label{relatedworks}
In recent years, RFF-based device authentication has gained popularity, and these approaches rely on machine learning techniques to extract the unique features of each wireless device. Huang et al. \cite{10433427} used density trace plots obtained from I/Q samples to authenticate wired and wireless devices. The three different deep learning methods, including 2D-CNN, the combination of 2D-CNN and Long Short-Term Memory (LSTM), and 3D-CNN, are used to extract features from each device. They verified the proposed approach by using I/Q data from five different ADALM PLUTO SDRs. Although they achieved an accuracy of 96.7\%, their work was focused on classifying devices and did not consider rogue device detection. Another work by Tong et al. \cite{jian2020deep} conducted a massive experimental study using data from over 10,000 WiFi and Automatic Dependent Surveillance-Broadcast (ADS-B) devices. They used a custom baseline model and a modified ResNet-50-1D as a machine learning model for feature extraction. Although the proposed technique achieved good performance in classifying devices, it was unable to detect rogue devices. Recent works have focused on utilizing a Siamese network for RFF to detect rogue devices. Siamese networks are trained to differentiate between similar pairs and dissimilar pairs.
For example, G. Sun et al. \cite{9661271} used a combined Siamese network to differentiate between eight genuine devices and three rogue devices. Similarly, Birnbach et al. \cite{birnbach2023adaptable} proposed using a Siamese network and verified their approach by employing both wireless (ADS-B) and wired (RS-485) communication links. 
In \cite{8879545}, Roy et al. proposed the Radio Frequency Adversarial Learning (RFAL) framework, which generates the synthetic rogue samples by using GANs and uses the discriminator of a trained GAN to differentiate between genuine devices and rogue devices.
Despite achieving significant performance in identifying genuine devices and detecting rogue devices, their work did not consider data from any real devices as rogue devices. Only synthetic samples generated by the generator of the GAN were considered as data from the rogue device. 

From a comprehensive review of these existing works, we found that several previous studies focus on detecting rogue devices. Some of them consider only real devices as rogue devices, while others focus solely on synthetically generated samples as rogue devices. However, in a real-world setting, an adversary can take any form; they could either use a real rogue device or attempt to mimic the characteristics of genuine devices. Therefore, our work considers both types of adversaries in the attack scenario.

\section{Methodology}\label{methodology}
\subsection{Data Collection}
Our data collection process involves the 11 different ADALM-PLUTO Software Defined Radios (SDRs) connected to the MATLAB environment. One ADALM-PLUTO is the receiver to receive the In-phase (I) and Quadrature (Q) signal components from the remaining 10 SDRs. The transmitter transmits real-time Orthogonal Frequency Division Multiplexing (OFDM) wireless communication signals using MATLAB with Communication Toolbox support for PLUTO SDR. The transmission systems are connected to two computers in a MATLAB environment configured with Communication Toolbox support for PLUTO SDR. All data collection was conducted in our indoor lab environment. 
From each device, we collected 19,920 frames, each consisting of a header with 72 I/Q samples and a payload consisting of 1,728 I/Q samples. Our work focuses solely on the header part, as it is suitable for extracting hardware-specific imperfections. Further detail of the data collection process is discussed in our previous article \cite{iot6030047}. 

\subsection{System Model}\label{systemmodel}

\begin{figure*}
\centering
\includegraphics[width=0.85\linewidth]{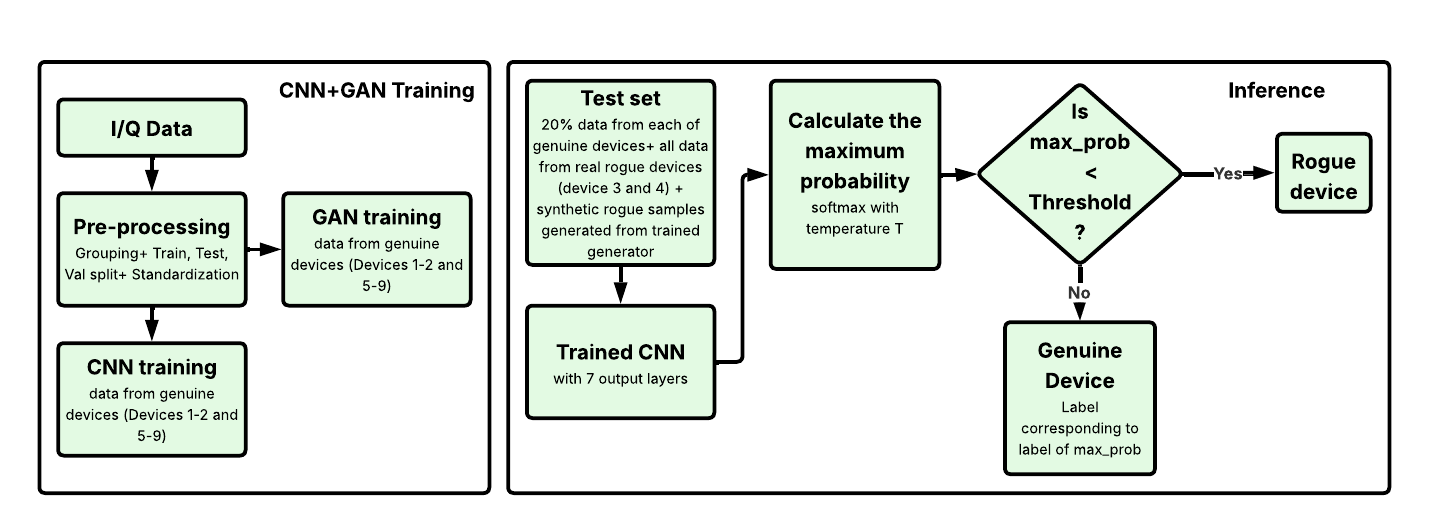}
\caption{RF Fingerprinting Pipeline using CNN and GAN } 
\label{pipeline}
\end{figure*}
The figure \ref{pipeline} shows the RF fingerprinting pipeline using CNN and GAN. We can observe two stages in our system, i.e., CNN and GAN training, and the inference stage. Also, the training stage starts with the preprocessing of I/Q data, where the data are merged, split into training, testing, and validation sets, and standardized. Out of our 10 devices, we consider seven random devices as genuine devices (devices 1-2 and 5-9), and two random devices (3 and 4) are chosen as rogue devices. The remaining device (10) is only used in the validation set. First of all, data from all devices is merged in such a way that 10 consecutive frames are combined to make a single frame, which means that, for each device that originally had 19920 frames with 72 I/Q samples in each frame, after merging, each device has 1992 frames with 720 I/Q samples in each frame. After merging, we create longer and more informative frames, each of which captures a wider range of temporal information.  Our training set comprises 70\% of data from the genuine devices, i.e., devices 1-2 and devices 5-9. The training set does not include I/Q samples from the remaining three devices. Our validation set consists of 10\% of data from genuine devices and all samples from the validation-only device, i.e., device 10. Remaining samples, i.e., 20\% data from genuine devices and all the data from the rogue devices, are used in the test set (where 1000 synthetic samples generated by a trained generator will be added later). This results in the training set having 9760 frames from all genuine devices, the validation set having 3386 samples from both genuine and rogue devices, and the test set having 7774 frames from genuine devices, rogue devices, and synthetically generated from the generator of GAN. Note that frames now have 720 I/Q pairs. 
\par 
After merging and splitting of I/Q data, we standardize the I/Q data using the mean and the standard deviation computed from the training set. This makes sure that both the I and Q components of the signal have zero mean and unit variance.

After standardization, the I and Q components are stacked together to form the final input with shape \((720, 2, 1)\), where 720 is the number of I/Q samples, 2 represents the in-phase and quadrature channels, and 1 is the channel dimension used for CNN input. The values of mean and standard deviation calculated from the training set are applied to the validation and test sets to avoid data leakage and maintain consistency.
The next step is to train the proposed CNN and GAN model using the training set, which consists of data from genuine devices only. The architectures of CNN and GAN are explained in sections \ref{cnn} and \ref{gan}, respectively. Here, the role of CNN is to detect the rogue and genuine devices and identify the frames detected as genuine devices. On the other hand, the role of GAN is to generate synthetic rogue samples, enabling us to address the fact that adversaries may not always be real rogue devices; sometimes, they may try to mimic the signal characteristics of genuine devices. 
\par
In the inference stage, the extended test set consisting of data from the rogue device, the genuine device, and synthetic rogue data generated by the trained generator is passed through the trained CNN. First, CNN differentiates the data, whether it is from genuine devices or from rogue devices. The output of CNN is the seven different probabilities, i.e., the vector of logits \( \mathbf{z} \in \mathbb{R}^7 \) corresponding to seven different classes of seven genuine devices. Then, these logits are passed through a softmax function with temperature scaling, generating class probabilities.  
\begin{equation}
\mathbf{p} = \text{softmax}\left( \frac{\mathbf{z}}{T} \right)
\end{equation}
where \( T \) is the temperature parameter used to control the confidence of the distribution. For each input sample, the maximum softmax probability is calculated as:
\begin{equation}
p_{\text{max}} = \max_{i} p_i
\end{equation}
This maximum probability value is compared with the predefined threshold. If the value is greater than the threshold, it is detected as a genuine device; otherwise, it is detected as a rogue device. 
\[
\hat{y} =
\begin{cases}
-1 & \text{if } p_{\text{max}} < \text{threshold} \quad \text{(classified as rogue)} \\
\arg\max_{i} p_i & \text{otherwise (classified as genuine device } i \text{)}
\end{cases}
\]
The data points that are detected as genuine devices are assigned to one of the classes of genuine devices based on the highest probability. This approach makes the model capable of performing open-set detection and closed-set classification. 
\subsection{CNN Architecture}\label{cnn}
The CNN model we are using here is responsible for differentiating between genuine and rogue devices and classifying detected genuine samples as explained in section \ref{systemmodel}. The input to the CNN is of the shape \((720, 2, 1)\). The CNN consists of one input layer, followed by one convolutional layer, one dense layer, and an output layer with softmax activation. The first convolutional layer is a two-dimensional convolutional layer having 32 filters and a kernel of size \((5, 1)\) with ReLU activation and L2 regularization, followed by batch normalization and maxpool of size \((2, 1)\). The result now is flattened and given to the dense layer with 352 neurons. The dense layer is equipped with the ReLU activation function and L2 regularization, and a dropout of 0.3. Finally, the output layer is the dense layer with seven neurons corresponding to seven genuine device classes. The model is compiled with the Adalm optimizer with a learning rate of 0.00036 and trained for 10 epochs with sparse categorical cross-entropy. 
\subsection{ Threshold Calculation}\label{thresholding}
The validation set with real genuine and real rogue devices is used to determine the threshold. We pass the validation set through the trained CNN, which then generates the softmax probabilities for each data point in the validation set. Then, the generated softmax probabilities are temperature-scaled, and the maximum softmax probability is chosen from each output vector. To find the best threshold, we select 100 different candidate threshold values at equal intervals between the minimum and maximum chosen probabilities. For each candidate threshold, \( \theta \), we apply the decision rule, i.e., if \( p_{\text{max}} < \theta \), the input is predicted as a rogue device; otherwise, it is predicted as the genuine device. For each candidate threshold, we calculate the f1-score of rogue devices, considering the detection problem as binary classification. The candidate threshold value that gives the highest value of the f1-score is selected as the best threshold \( \theta^\ast \). 
\subsection{Hyperparameter Tuning of CNN}
We performed the hyperparameter tuning using \texttt{KerasTuner}'s random search method to optimize the performance of the proposed CNN model. The tuning process involves searching over a defined hyperparameter space to identify the combination of parameters that gives the best validation accuracy. The total number of trials for different combinations of hyperparameters is 20. The table \ref{tab:hyperparam_space} illustrates the hyperparameters with their respective ranges used during different trials. The key parameters include the number of convolutional layers, number of filters, kernel sizes, number of dense layers, dense units, dropout rates, learning rate, and L2 regularization strength.
\begin{table}[h!]
\centering
\caption{Hyperparameter Search Space for CNN Tuning}
\label{tab:hyperparam_space}
\begin{tabular}{|p{3.15cm}|p{4.9cm}|}
\hline
\textbf{Hyperparameter} & \textbf{Search Range / Values} \\
\hline
Number of Conv Layers & \{1, 2, 3\} \\
\hline
Filters per Conv Layer & \{32, 64, 96, 128\} \\
\hline
Kernel Size per Conv Layer & \{3, 5, 7\} \\
\hline
L2 Regularization & \( [10^{-5}, 10^{-2}] \) (log scale) \\
\hline
Number of Dense Layers & \{1, 2\} \\
\hline
Dense Units per Layer & \{32, 96, 160, 224, 288, 352, 416, 480, 544\}\\
\hline
Dropout Rate & \([0.3, 0.7]\) (step = 0.1) \\
\hline
Learning Rate (Adam) & \( [10^{-4}, 10^{-2}] \) (log scale) \\
\hline
Batch Size & 128 (fixed) \\
\hline
Epochs & 10 (fixed) \\
\hline
\end{tabular}
\end{table}
The top five models are identified from the hyperparameter tuning, and we apply the different values of temperature scaling, i.e., \( T = \{1.0, 1.5, 2.0, 2.5, 3.0\} \), to the model to identify the best value of \( T \). In this process, the same thresholding approach as described in section \ref{thresholding} is used to select the best threshold in each particular instance of \(T\). Here, the goal is to select the value of T that gives the maximum value of the f1-score of the rogue class. In doing this, the best value for T was obtained to be 2.5. 
The best-performing model achieved an f1-score of 0.9871 for rogue detection at the software temperature of 2.5 and the threshold of 0.1987. After hyperparameter tuning, we obtained the best CNN model as described in section \ref{cnn}. 
\subsection{GAN Architecture}\label{gan}
The GAN consists of two models, i.e., a generator ($G$) and a discriminator ($D$). The generator generates fake samples from the given data distribution, and the discriminator calculates the similarity between generated fake samples and real data samples. 
\begin{itemize}
    \item \textbf{Generator ($G$):}
    The generator takes input of random noise $n$. Initially, signals generated by the generator are random, and the generated samples improve over time, learning from the discriminator and resembling the real trusted data. The random noise vector input given to the generator passes through three dense layers with 2048, 4096, and 1440 neurons, respectively. After the first two layers, batch normalization and ReLU activation are applied. Then, the output of the final layer is reshaped to produce an I/Q sample of (720, 2). 
    \item \textbf{Discriminator ($D$):} 
    The synthetic samples generated by the generator and the trusted samples from real devices are the input to the discriminator $D$. The discriminator learns to differentiate between trusted and synthetic samples. The main goal is to train $G$ so that the probability of making a mistake $D$ is maximized. The output of the discriminator is fed back to the generator to tune its hyperparameters. The discriminator takes the input of shape (720, 2) and passes it through two one-dimensional convolutional layers with 64 and 128 filters, with kernel sizes of 7 and 5. Both of the convolutional layers are followed by LeakyReLU activation ($\alpha$ = 0.2) and dropout (0.3). After this, the output layer is flattened, and the output is passed through the dense layer with 128 neurons and the LeakyReLU activation function. A single logit is produced by the final dense layer predicting real or fake samples. The Discriminator is trained using mean squared error loss. In contrast, the generator is trained using feature matching loss, depending on the distance between the average discriminator feature of the real and generated samples. 
\end{itemize}
Both the generator and discriminator are trained with the Adam optimizer with a learning rate of 0.001, and they are trained for 3000 epochs. All the model hyperparameters are inspired by the GAN used in Roy et al. \cite{8879545}. 

\section{Results} \label{results}
\subsection{F1-score vs Threshold}\label{f1scorevsthresholdres}
As explained in section \ref{thresholding}, we use a validation set to determine the threshold for the trained CNN to differentiate between genuine and rogue devices. For each sample in the validation set, the CNN yields a temperature-scaled softmax probability vector consisting of seven different probabilities, corresponding to each of the genuine devices. Then, the maximum of each of the vectors is calculated, and we will have a probability for each of the samples. From the maximum to the minimum of these probabilities, we choose 100 different candidate thresholds, and each of these thresholds is used to differentiate between genuine and rogue classes. For each of the candidate thresholds, we calculate the f1-score for the rogue class. 
The threshold candidate that gives the highest f1-score for the rogue class is selected as the best threshold \( \theta^\ast \). In our case, the best threshold was found to be \( \theta^\ast = 0.1987 \). 

\begin{figure}
    \centering
    \includegraphics[width=0.85\linewidth]{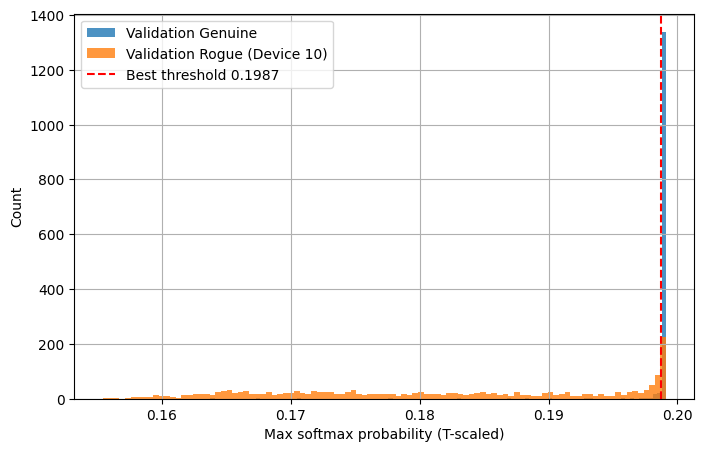}
    \caption{Distribution of max softmax probabilities for validation set}
    \label{distribution}
\end{figure}

The figure \ref{distribution} shows the distribution of maximum temperature-scaled softmax probabilities for the validation set consisting of both genuine and rogue devices. The blue-colored bins in the figure represent the genuine devices, and they are tightly clustered close to the higher end of the probability range. These tight clusters at the higher end of the probability range indicate that the model has strong confidence in genuine classes. On the other hand, the orange colors in the diagram are used to represent the rogue device (Device 10). The bins for the rogue devices are spread in the broader range within the comparatively low probability regions. The red dashed line is the threshold determined in section \ref{f1scorevsthresholdres}. This threshold line effectively separates the genuine and rogue devices. This separation highlights that the trained CNN model generates lower probabilities for the rogue devices, making it a suitable metric to differentiate rogue and genuine devices.
\subsection{Comparison of Real and Generated I/Q Constellations}
\begin{figure}
    \centering
    \includegraphics[width=0.85\linewidth]{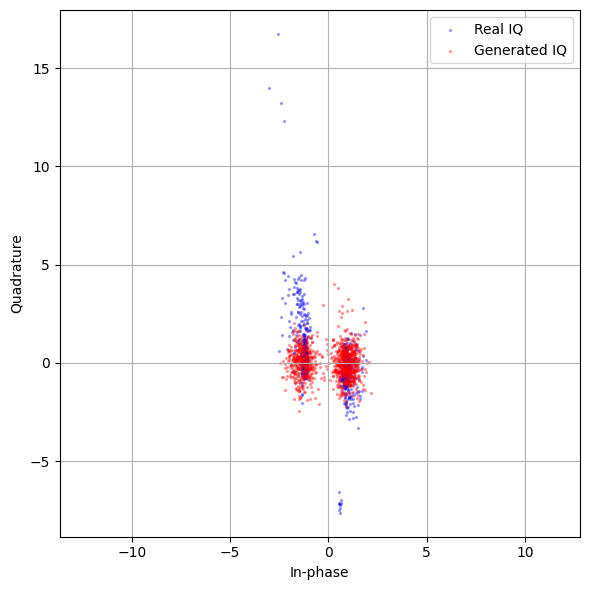}
    \caption{Real vs generated I/Q constellation}
    \label{constellationcomparision}
\end{figure}

We trained the GAN to generate synthetic rogue samples using its trained generator. As discussed earlier in section \ref{gan}, 1000 synthetic rogue samples were generated and appended to the initial test set, which already included real rogue and genuine device samples. The purpose of including synthetic rogue samples in the test set is to emulate an adversarial attack scenario, where adversaries may not always use real rogue devices. Instead, they might attempt to mimic the characteristics of genuine devices to gain unauthorized access to the system. To validate that the synthetic samples generated by the generator resemble the real I/Q samples, or genuine samples from the training set, we plotted the I/Q constellation of 1000 I/Q pairs from each of the generated synthetic and real genuine samples, as shown in Figure \ref{constellationcomparision}. In the figure, real I/Q samples are shown in blue and synthetic I/Q samples in red. We can observe that the generated samples follow the distribution of real signals closely, forming clusters in similar regions of the I/Q plane. Although there are minor deviations, the figure visually demonstrates that the generator is capable of mimicking the RF signal characteristics of real genuine devices.
Furthermore, we used a widely used statistical metric \textit{Fréchet Distance (FD)} to compare the similarity between real and generated samples. We generated the same number of samples as in the training set and compared them with the real samples in the training set. The FD calculation process models the real and generated samples as a multivariate Gaussian distribution by calculating their empirical mean and covariance matrices. The Fréchet Distance between the two distributions is given by \cite{dowson1982frechet} as follows:
\begin{equation}
\text{FD} = \|\mu_r - \mu_g\|^2 + \mathrm{Tr}\left(\Sigma_r + \Sigma_g - 2\left(\Sigma_r \Sigma_g\right)^{\frac{1}{2}}\right)
\end{equation}
where: $\mu_r$, $\Sigma_r$ is the mean vector and covariance matrix of the real I/Q distribution, $\mu_g$, $\Sigma_g$ is the mean vector and covariance matrix of the generated I/Q distribution, $\mathrm{Tr}(\cdot)$ is the trace operator, and $\left(\Sigma_r \Sigma_g\right)^{1/2}$ is the matrix square root of the product of the two covariances.

With this formula, we calculated the FD score between real and generated samples to be 0.0545. The low value of the FD score indicates that the real and generated sample distributions are close to each other, validating the generator's ability to generate synthetic samples. 
\subsection{Confusion matrices demonstrating the performance of the proposed approach}
\begin{figure}
    \centering
    \includegraphics[width=0.75\linewidth]{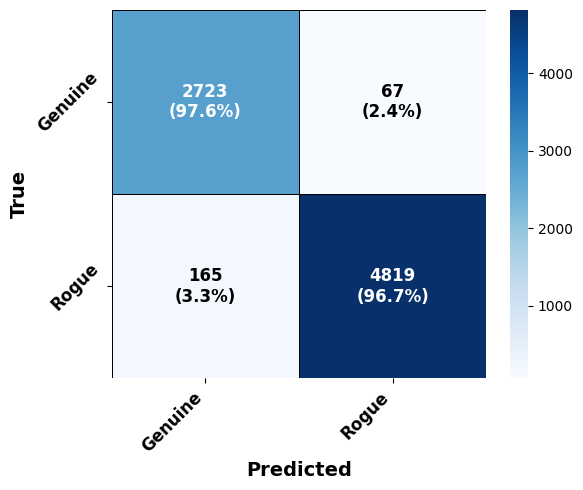}
    \caption{Performance of the proposed approach to distinguish between rogue and genuine devices.}
    \label{Rogue vs Genuine}
\end{figure}

\begin{figure}
    \centering
    \includegraphics[width=0.85\linewidth]{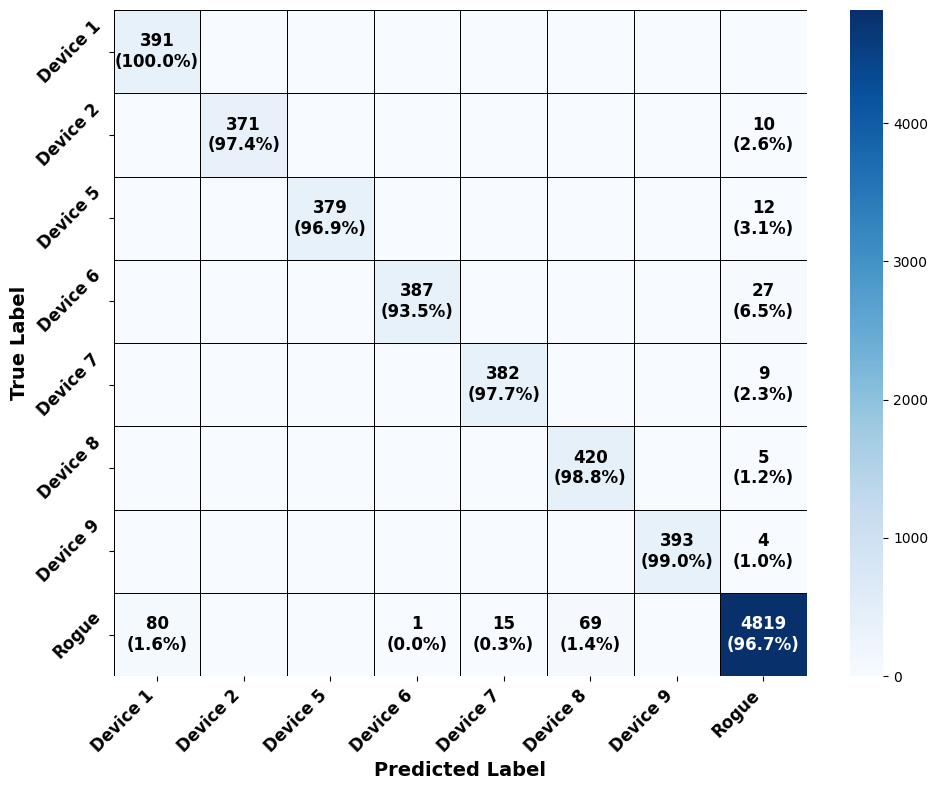}
    \caption{ Overall confusion matrix showing both the detection of genuine and rogue devices and the classification of genuine devices into their respective classes.}
    \label{overall}
\end{figure}

The test sets, consisting of 2790 genuine samples from seven genuine devices, 3984 samples from rogue devices (devices 3 and 4), and 1000 synthetic samples generated by the generator, are used to evaluate the performance of the proposed method. At first, this test set is passed through a trained CNN to differentiate between genuine and rogue devices by using temperature-scaled softmax probability thresholding. The figure \ref{Rogue vs Genuine} shows the performance of the proposed method in differentiating genuine and rogue samples. We can observe that 97.6\% of genuine samples are classified as genuine, and 96.7\% of rogue samples are detected as rogue devices during the binary classification. Also, 3.3\% of rogue samples are classified as genuine, and 2.4\% of genuine samples are classified as rogue devices. After this binary classification, the samples classified as genuine samples are labelled with one of the seven genuine devices. After doing so, we obtained the performance of our system as shown in figure \ref{overall}. We can observe that all samples detected as genuine devices are classified into their correct corresponding labels.   The classification accuracy among genuine devices is high across all classes, with most devices achieving above 97\%, and several devices, such as device 1 and device 9, achieving near-perfect or perfect classification. Misclassification of rogue samples into genuine classes is minimal, with only a small portion being incorrectly predicted as genuine. These results demonstrate that the proposed CNN-based model not only effectively detects both real and synthetic rogue devices but also accurately classifies genuine devices into their respective identities once they pass the softmax confidence threshold.
\section{Limitation and Future works}\label{lim}
Although the proposed approach performs promisingly in detecting rogue devices and identifying genuine ones, this work still has some limitations that can be addressed in future research. First, our study is limited to data collected from ten devices, whereas real-world wireless, IoT, and UAV networks may involve hundreds of devices. Future work can incorporate more devices during data collection to evaluate the scalability of the proposed approach. Second, all data were collected from stationary devices. However, real-time systems often involve mobile nodes, and device movement may affect the performance of the proposed method. This limitation can be addressed by incorporating mobile nodes into the future data collection process.
\section{conclusion} \label{conc}
This paper presents a CNN-based method for detecting genuine and rogue devices (including GAN-generated) using RFF and softmax thresholding. A GAN was trained to generate fake I/Q samples that resemble genuine I/Q samples, simulating an adversarial scenario in which an attacker attempts to mimic the RF fingerprints of genuine devices. To evaluate the model's effectiveness, we used the I/Q data collected from ten Adalm Pluto SDRs and GAN-generated adversarial samples. Although the proposed approach was verified using I/Q data from a limited number of devices, i.e., 10 devices, it achieved a detection accuracy of 96.7\% for rogue devices and 97.6\% for genuine devices. In particular, all genuine devices were correctly identified and accurately classified. These results demonstrate the effectiveness of the proposed method in detecting both genuine and rogue devices, including adversaries that attempt to mimic the RF characteristics of genuine wireless devices.

\end{document}